\documentclass[aps,prl,twocolumn,superscriptaddress,showpacs]{revtex4-1}

\usepackage{graphicx}
\usepackage{dcolumn} 
\usepackage{bm}      
\usepackage{color}

\usepackage{amsmath,amssymb}

\usepackage{epstopdf}
\usepackage{latexsym}
\usepackage{subfigure}
\usepackage{dsfont} 
\usepackage{wasysym} 
\usepackage{hyperref} 
\usepackage{times}
\hypersetup{
	colorlinks   = true,
	citecolor    = blue,
	linkcolor    = blue,
	urlcolor     = blue
}

\begin{document}
	
	\title{Removing the spin ice cap: magnetic ground states of rare earth tripod kagome lattice  Mg$_2$RE$_3$Sb$_3$O$_{14}$ (RE = Gd, Dy, Er)}
	
	\author{Z.~L.~Dun}
	\affiliation{Department of Physics and Astronomy, University of Tennessee, Knoxville, Tennessee 37996-1200, USA}
	
	\author{J. Trinh}
	\affiliation{Department of Physics, University of California, Santa Cruz, CA 95064, USA}
	
	\author{K. Li}
	\affiliation{Beijing National Laboratory for Molecular Sciences, State Key Laboratory of Rare Earth Materials Chemistry and Applications, College of Chemistry and Molecular Engineering, Peking University, Beijing 100871, PR China}
	\affiliation{Center for High Pressure Science and Technology Advanced Research, Beijing, 100094, PR China}
	
	\author{M.~Lee}
	\affiliation{Department of Physics, Florida State University, Tallahassee, FL 32306-3016, USA}
	\affiliation{National High Magnetic Field Laboratory, Florida State University, Tallahassee, FL 32310-3706, USA}
	
	\author{K. W. Chen}
	\affiliation{National High Magnetic Field Laboratory, Florida State University, Tallahassee, FL 32310-3706, USA}
	
	\author{R. Baumbach}
	\affiliation{National High Magnetic Field Laboratory, Florida State University, Tallahassee, FL 32310-3706, USA}
	
	\author{Y. F. Hu}
	\affiliation{Beijing National Laboratory for Molecular Sciences, State Key Laboratory of Rare Earth Materials Chemistry and Applications, College of Chemistry and Molecular Engineering, Peking University, Beijing 100871, PR China}
	
	\author{Y. X. Wang}
	\affiliation{Beijing National Laboratory for Molecular Sciences, State Key Laboratory of Rare Earth Materials Chemistry and Applications, College of Chemistry and Molecular Engineering, Peking University, Beijing 100871, PR China}
	
	\author{E.~S.~Choi}
	\affiliation{National High Magnetic Field Laboratory, Florida State University, Tallahassee, FL 32310-3706, USA}
	
	\author{B. S. Shastry}
	\affiliation{Department of Physics, University of California, Santa Cruz, CA 95064, USA}

	\author{A. P. Ramirez}
	\affiliation{Department of Physics, University of California, Santa Cruz, CA 95064, USA}

	\author{H.~D.~Zhou}
	\affiliation{Department of Physics and Astronomy, University of Tennessee, Knoxville, Tennessee 37996-1200, USA}
	\affiliation{National High Magnetic Field Laboratory, Florida State University, Tallahassee, FL 32310-3706, USA}
	
	\date{\today}
	
	\begin{abstract}
		We present the structural and magnetic properties of a new compound family, Mg$_2$RE$_3$Sb$_3$O$_{14}$ (RE = Gd, Dy, Er), with a hitherto unstudied frustrating lattice, the ``tripod kagome" structure. Susceptibility (ac, dc) and specific heat exhibit features that are understood within a simple Luttinger-Tisza type theory. For RE = Gd, we found long ranged order (LRO) at 1.65 K, which is consistent with a 120 $^{\circ}$ structure, demonstrating the importance of diople interactions for this 2D Heisenberg system. For RE = Dy, LRO at 0.37 K is related to the ``kagome spin ice (KSI)" physics for a 2D system. This result shows that the tripod kagome structure accelerates the transition to LRO predicted for the related pyrochlore systems.  For RE = Er, two transitions, at 80 mK and 2.1 K are observed, suggesting the importance of quantum fluctuations for this putative XY system.  
	\end{abstract}
	\maketitle
	
	\textit{Introduction.---}%
	The two-dimensional (2D) kagome lattice magnet (KLM) has been a favorite in the theoretical condensed matter community since the experimental work on SCGO \cite{SCGO}, due to the strong frustration associated with its network of corner-shared triangles. Many exotic states are predicted, such as the quantum spin liquid (QSL) state \cite{Frustration2,Wen,NatureLB}, the spin-orbital liquid state \cite{SOL}, the kagome spin ice (KSI) state \cite{PRB2002Wills}, dipolar spin order \cite{PRB2015Dipoles}, the Kosterlitz-Thouless (KT) phase transition \cite{PRBKT}, Quantum Order by Disorder (QObD) \cite{Frustration1}, and nematicity and supernematicity \cite{Nematic}. 
	The large variety of exotic states predicted for kagome spin systems lies in contrast to a paucity of experimental systems. Early efforts include the exploration of langasites RE$_3$Ga$_5$SiO$_{14}$\cite{Lang1,Lang2,Lang3}, which possess a distorted kagome lattice. Recent attention has been paid to vesignieite BaCu$_3$V$_2$O$_8$(OH)$_2$ \cite{BaCuV} and herbertsmithite ZnCu$_3$(OH)$_6$Cl$_2$ \cite{Nature2012}. The later one shows intriguing signs of QSL behavior \cite{Nature2012}.  From a materials standpoint, however, these two systems are limited by (i) known defect prone structures \cite{BaCuV,Sitedisorder}, and (ii) the inability to substitute facilely on the magnetic site (e.g with non-Heisenberg spins) to realize states other than the QSL.  Clearly then, finding new KLM-containing compounds with spin-type variability is a challenge of the highest order.

	Intriguingly, a 2D KLM is naturally contained in the frustrated 3D pyrochlore structure. In pyrochlores RE$_2$X$_2$O$_7$ (RE = rare earth element, X = transition metal element), both the RE$^{3+}$ and X$^{4+}$ sublattices form alternating kagome and triangular layers along the [111] axis as a result of the corner-shared tetrahedrons (Fig. 1(a)) \cite{Tabata}. However, the strong inter-layer interaction enforces three-dimensionality.  An exception is found in studies of Dy$_2$Ti$_2$O$_7$ in a [111] magnetic field, which polarizes the triangular layer spins, effectively decoupling the kagome planes, leading to a KSI state \cite{Tabata}. 
	
	Obviously, if one can remove the magnetic moment of the triangular layer in the pyrochlore lattice, a RE$^{3+}$ kagome-only lattice might be realized, enabling the study of intrinsic kagome physics. Because of various spin and spin anisotropies of different RE$^{3+}$ ions, exotic and rich magnetic properties should be immediately available via the complex interplay among the spin-orbital coupling, dipolar interaction, and exchange couplings. In pyrochlores, for example, this interplay leads to multi-$k$ ordering \cite{GdTi1} with multiple field induced transitions \cite{GdTi3} for Heisenberg spins in Gd$_2$Ti$_2$O$_7$, the spin ice state \cite{Science2001} for Ising spins in Ho$_2$Ti$_2$O$_7$ \cite{PRL1997} and Dy$_2$Ti$_2$O$_7$ \cite{Nature1999}, and QObD physics in the XY spin system Er$_2$Ti$_2$O$_7$ \cite{ErTiLB}. Then, what will be the magnetic ground states in the RE-based KLMs?
	
	In this letter, we have created such a KLM - Mg$_2$RE$_3$Sb$_3$O$_{14}$ based on partial ion substitution in the pyrochlore lattice.  Here, the triangular layers in the pyrochlore structure are occupied by non-magnetic Mg$^{2+}$ ions, leaving the RE$^{3+}$-kagome layer well isolated from neighboring layers.  We studied three representative systems (RE = Gd, Dy, Er)  by dc-, ac-susceptibility ($\chi_{dc}$, $\chi_{ac}$), and specific heat (C$(T)$) measurements. We present a spin Hamiltonian and show the fundamental differences in collective behavior between the 2D KLMs and their 3D pyrochore cousins.  
	
	\begin{figure*}[tp]
		\linespread{1}
		\par
		\begin{center}
			\includegraphics[width= 6.5 in]{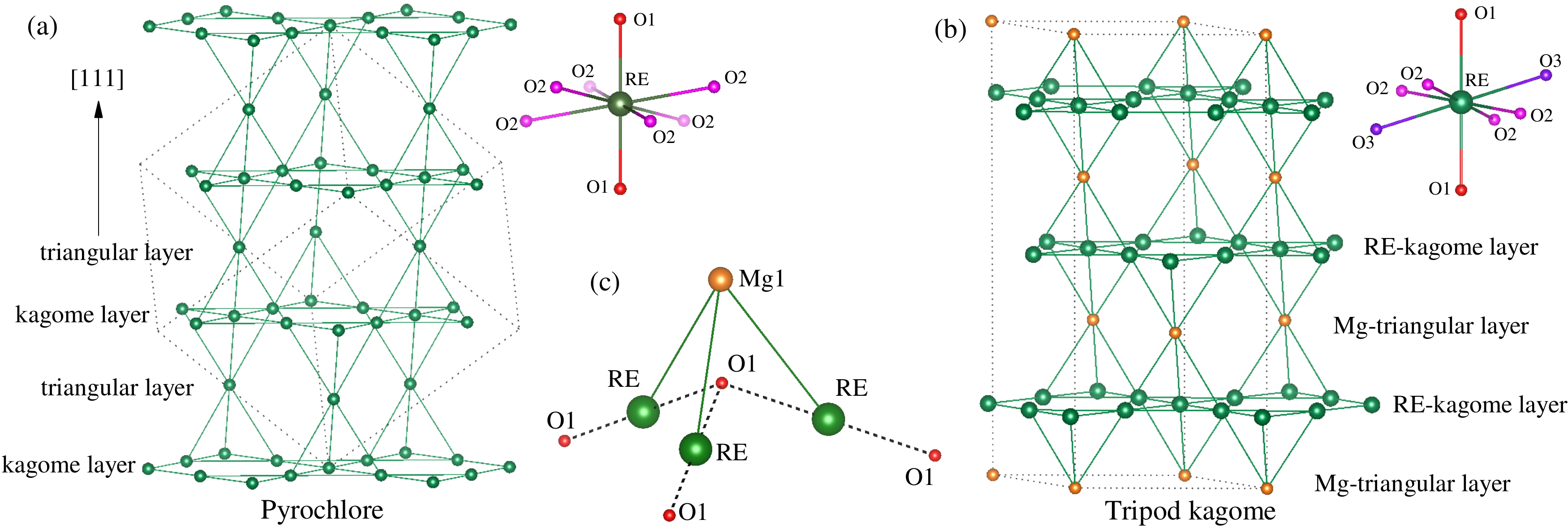}
		\end{center}
		\par
		\caption{\label{Fig:1}(color online) (a) Alternating kagome and triangular layers in a  pyrochlore lattice viewing along the [111] axis. Dashed lines indicate a single unit cell.  (b) Alternating RE$^{3+}$-kagome layers and Mg-triangular layers in a tripod kagome lattice. Local oxygen environment around RE$^{3+}$ ion for the two structure is shown at the top-right. (c) A single ``tripod". Dashed lines represent Ising axes.}
	\end{figure*}

	\textit{Structure.---}%
	Sample synthesis method and measurement setups are described in the supporting material \cite{SM}.  The Mg$_2$RE$_3$Sb$_3$O$_{14}$ (RE = Gd, Dy, Er) structure is rhombohedral with R-3m space group in hexagonal expression. Compared with the pyrochlore lattice,  the triangular layers of both the RE$^{3+}$ and Sb$^{5+}$ sublattices in the KLM structure are occupied by Mg$^{2+}$ (the Mg$^{2+}$ site can also be replaced by Co$^{2+}$ \cite{JSSC2014Co}, Mn$^{2+}$ \cite{JSSC2014Mn} and Zn$^{2+}$ \cite{Cava}). Thus the chemical formula can also be  written as (Mg$_{0.25}$RE$_{0.75}$)$_2$(Mg$_{0.25}$Sb$_{0.75}$)$_2$O$_{7}$, which is a pyrochlore (RE$_2$X$_2$O$_7$) structure with $\frac{1}{4}$ RE$^{3+}$ and X$^{4+}$ ions substituted in an ordered manner (Fig. 1(b)). It is noteworthy that for the X-ray diffraction pattern of Mg$_2$RE$_3$Sb$_3$O$_{14}$ \cite{SM}, the strongest peak for pyrochlore at 2$\theta$ $\sim$ 30$^{\circ}$ disappears completely and splits into two peaks, providing evidence for the absence of Mg-RE or Mg-Sb site-disorder \cite{JSSC2014Mn}. As shown below, the C$(T)$ peaks at their phase transitions are very sharp, further underscoring the high degree of site order in the kagome layers. This good kagome layer separation is likely due to the large ion size difference between Mg$^{2+}$ and RE$^{3+}$. In this structure, the nearest neighbor distance between the RE$^{3+}$ ions within a kagome layer remains similar to that of its pyrochlore cousin, and the RE$^{3+}$-kagome layers are well isolated from each other by the non-magnetic Mg$^{2+}$ and Sb$^{5+}$ layers. Take Mg$_2$Gd$_3$Sb$_3$O$_{14}$ for example, the nearest Gd-Gd distance within a kagome layer (3.678 \AA) is similar to that in Gd$_2$Ti$_2$O$_{7}$ (3.600 \AA), and much smaller than that between different planes (6.162 \AA). Since the dipole-dipole energy goes as 1/$r^3$, this leads to inter-layer energies an order of magnitude smaller than intra-layer energies. Thus, the kagome lattice in Mg$_2$RE$_3$Sb$_3$O$_{14}$ is seemingly free of structural defects.
	
	In RE$_2$X$_2$O$_7$, one important structural feature is that each RE$^{3+}$ ion is surrounded by eight oxygens (Fig. 1(a)) with two shorter RE-O1 bonds lying along the local-[111] axis and six longer RE-O2 bonds forming a puckered ring. This feature defines the crystal electric field (CEF) and the g-factor which determines the ionic anisotropy for the  RE$^{3+}$ spins. In Mg$_2$RE$_3$Sb$_3$O$_{14}$, this local oxygen coordination is largely preserved. The RE ion is still surrounded by eight oxygens with the two shortest RE-O1 bonds remain lying along the local-[111] axis (Fig. 1(b)). The difference is that the longer six RE-O bonds are divided into two sets: four longest RE-O2 bonds and two intermediate RE-O3 bonds \cite{SM}. Since the CEF degeneracy has already been lifted by the pyrochlore-like anisotropy for an effective spin-1/2 system, the dominant anisotropy remains the one distinguishing the puckered ring from the local-[111] oxygens, making this in-plane anisotropy most likely irrelevant for the ground state degeneracy. 
	
	Given the high degree of site order, the large difference in separation between intra-plane and inter-plane RE ions, it is appropriate to consider this a well-formed kagome structure.  In addition, the CEF-driven single-ion anisotropy, which is vestigial from the parent pyrochlore structure, defines directions for either the Ising spins or the XY-spin normal vectors that are neither uniaxial nor uniplanar.  This particular situation of three distinct axes with specific inter-axes angles will be important for understanding ordered spin configurations, as we show below.  Given the uniqueness of this structure and the need to distinguish it from KLMs with undefined local anisotropy, we call this the \textit{`` tripod kagome lattice (TKL)"}, inspired by a ``tripod" formed by three RE$^{3+}$ and one Mg$^{2+}$ ion (Fig. 1(c)).

	\begin{figure*} [tp]
		\linespread{1}
		\par
		\begin{center}
			\includegraphics[width= 6.7 in]{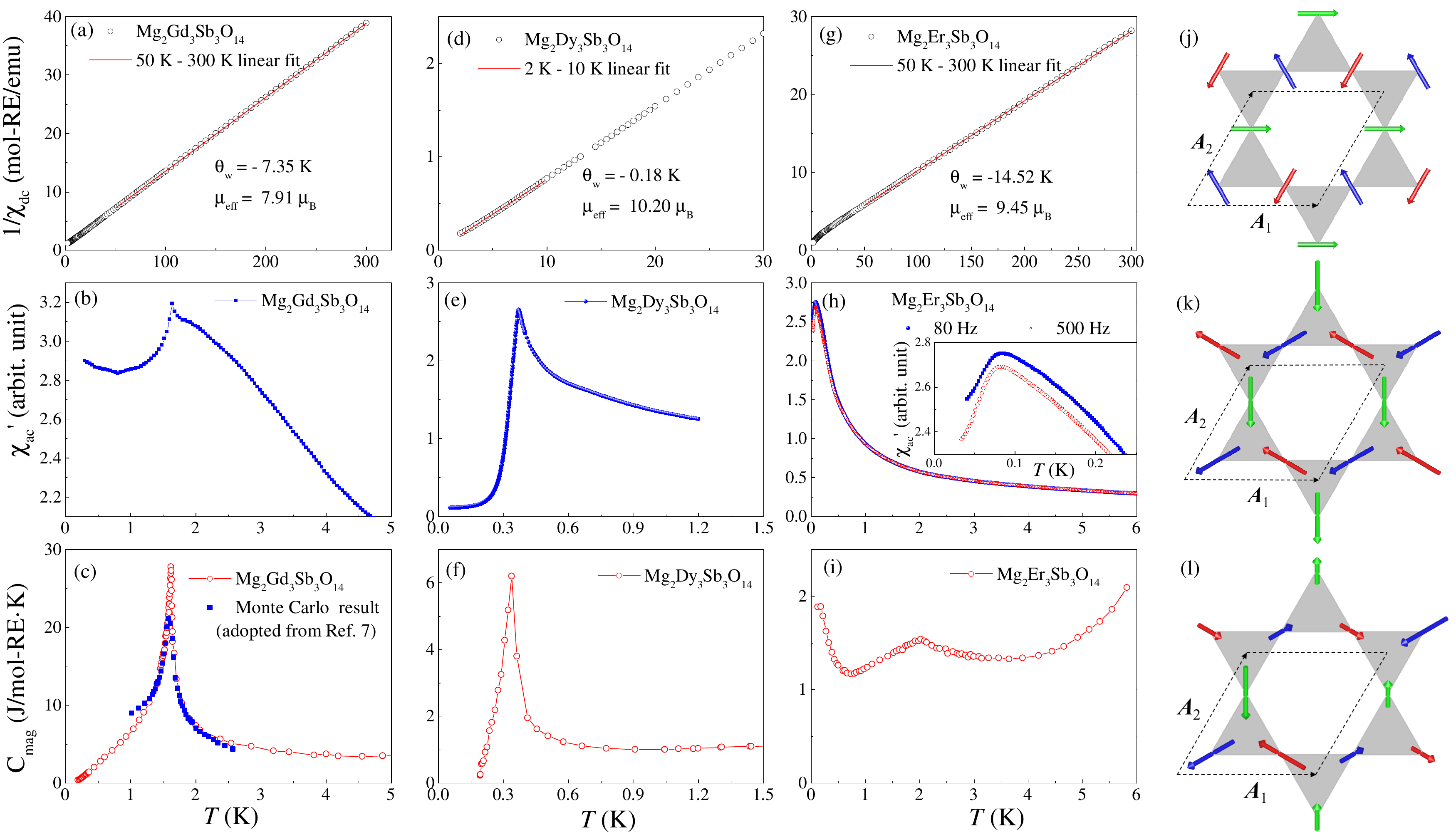}
		\end{center}
		\par
		\caption{\label{Fig:2}(color online) (a-i) Temperature dependence of the inverse $\chi_{dc}$ measured with 200 Oe, real part of the $\chi_{ac}$, and magnetic specific heat C$_{mag}$ for three compounds. The blue dot in (c) is the Monte Carlo simulation result adopted from Ref. \cite{PRB2015Dipoles} with scaling. (j) The 120$^{\circ}$ LRO state for Mg$_2$Gd$_3$Sb$_3$O$_{14}$  and Mg$_2$Er$_3$Sb$_3$O$_{14}$ . The dash lines represent a unit cell with ${\bf A_1}$ and  ${\bf A_2}$ as basis vectors. (k) A $k$ = 0 LRO state, and (l) a  $k$ = (1/3, 2/3) SDW-like state for Mg$_2$Dy$_3$Sb$_3$O$_{14}$.}
	\end{figure*}

	\textit{Magnetic properties.---}%
	For Mg$_2$Gd$_3$Sb$_3$O$_{14}$, a Curie-Weiss (CW) fit from 50 $\sim$ 300 K of  $1/\chi_{dc}$ (Fig. 2(a)) yields a Weiss temperature, $\theta_{W}$ = -7.35 K and an effective magnetic moment, $\mu_{eff}$ = 7.91 $\mu_{B}$. The negative $\theta_{W}$ value is close to that of  Gd$_2$Ti$_2$O$_7$ ($\theta_{W}$ = -11.7 K) \cite{JPCM2000}. The value for  $\mu_{B}$ is consistent with $\mu_{eff}$ = 7.94 $\mu_{B}$ expected for Gd$^{3+}$ ($\textsuperscript{8}S_{7/2}$). With measurement frequencies ranging from 80 to 700 Hz, $\chi_{ac}$ shows a sharp and frequency independent peak at T$_N$ = 1.65 K (Fig.2 (b)),  indicating a LRO transition.  This transition is further confirmed by a sharp peak at the same temperature in magnetic specific heat $C_{mag}(T)$ (Fig. 2(c)). The magnetic entropy  between 0.2 and 6 K is 17.16 J/mol-Gd$\cdot$K \cite{SM}. This value is close to Rln(2S+1) = 17.29 J/mol$\cdot$K for a S = 7/2, indicating a complete LRO among the Gd$^{3+}$ spins.

	For Mg$_2$Dy$_3$Sb$_3$O$_{14}$, the CW fit below 10 K yields  $\theta_{W}$ = -0.18 K and  $\mu_{eff}$ = 10.20 $\mu_{B}$ (Fig. 2(d)), consistent with the free ion moment of  10.63 $\mu_{B}$ for Dy$^{3+}$ ($\textsuperscript{6}H_{15/2}$). In Dy$_2$Ti$_2$O$_7$ \cite{JPCM2000}, the small negative $\theta_{W}$ (-0.20 K) is due to competition between the dipolar interaction and super-exchange couplings of Dy ions.  Here, the similarity in local structure translates into similar-size coupling to the pyrochlore case, since the total spin-spin coupling is dominated by the dipolar interaction.  With the ferromagnetic dipolar interaction, the negative $\theta_{W}$  again shows the antiferromagnetic(AFM) nature of the Dy$^{3+}$-Dy$^{3+}$ exchange interactions in Mg$_2$Dy$_3$Sb$_3$O$_{14}$.
	A transition to LRO at $T_N$ = 0.37 K is observed both  in the $\chi_{ac}$ (Fig. 2(e)) and $C_{mag}(T)$ (Fig. 2(f)). The integrated magnetic entropy below 6 K is 5.38  J/mol-Dy$\cdot$K \cite{SM}, which is close to Rln2 = 5.76 J/mol-Dy$\cdot$K, as expected for a Kramers-doublet. This suggests the Dy$^{3+}$  spins fully order below 0.37 K.

	For Mg$_2$Er$_3$Sb$_3$O$_{14}$, the CW fit above 50 K yields $\theta_{W}$ = -14.25 K, and $\mu_{eff}$ = 9.45 $\mu_{B}$ (Fig. 2(g)), consistent with the free ion moment of $\mu$ = 9.58 $\mu_{B}$ for Er$^{3+}$ ($\textsuperscript{4}I_{15/2}$).  The value for $\theta_{W}$ is close to that of the pyrochlore Er$_{2}$Ti$_{2}$O$_{7}$ ( $\theta_{W}$ = -15.93 K \cite{JPCM2000}). The  $\chi_{ac}$ was measured down to 30 mK with a broad peak observed around 80 mK (Fig. 2 (i)), which shows weak frequency dependence.  The $C_{mag}(T)$ was measured down to 120 mK and exhibits a weak and broad peak around 2 K (Fig. 2(i)). At this temperature, no anomaly is observed in $\chi_{ac}$, while an extremely weak anomaly (2*10$^{-8}$ emu/mole-Er) was seen in  $\chi_{dc}$ at 2.1 K that is perhaps related to the weak $C_{mag}(T)$ peak.

	\textit{Theoretical investigation and Discussion.---}%
	The three systems discussed here - Gd, Dy, and Er, are likely representatives of the three different spin types, Heisenberg, Ising and XY, respectively, evidenced by similar low temperature magnetization curves compared with their pyrochlore cousins \cite{SM}.  In the 3D pyrochlore systems discussed above, each spin type yields significantly different behavior.  To uncover the possible differences among the spin types in the TKLs, we have used a Luttinger-Tisza type theory \cite{LT1, LT2} and studied the eigenvalues and eigenfunctions of the interaction matrix in wave vector space. We construct a 2D kagome lattice with ${\bf A_1}$ and ${\bf A_2}$ as basis vectors of the triangular Bravais lattice where there are three basis sites in a unit cell for an upright triangle of basis spins, labeled as blue, red and green (Fig. 2(j)). Thus the general Hamiltonian for the TKL can be written as \cite{SM}:
	\begin{equation}
	\label{eq:H}
	H=\frac{1}{2}\sum_{k,\alpha,\beta,a,b} S^{\alpha,a}\left(k\right)S^{\beta,b}\left(-k \right)V_{ab}^{\alpha\beta}\left (k\right) 
	\end{equation}
	where $V$ is the sum of a dipolar part, exchange part and a single ion anisotropy part dictated by the CEF effects. Here $\alpha$ and $\beta$ are the cartesian indices of the spins and a,b run over the three basis sites in unit-cell. The spin vector is the Fourier component of the real space object, and $k$ runs over the Brillouin zone of the triangular lattice. Thus for a given value of $k$, $V$ is a 9$\times$9 matrix that can be easily diagonalized. The dipolar part ($D_{nn}$) is fixed exactly by the effective moment of spin and distances between RE ions, while the exchange ($J_{ex}$) and single ion terms are found from the $\theta_{W}$ and the CEF splitting of the RE$^{3+}$ in the given environment \cite{SM}.
	
	The Mg$_2$Gd$_3$Sb$_3$O$_{14}$ has Heisenberg spins with $J$ = 7/2, $L$ = 0 and therefore no single ion term. The $\theta_{W}$ of -7.35 K leads to an estimate of the exchange constant $J_{ex}$ $\sim$ 6.10 K, while the dipolar energy scale of nearest neighbor spins $D_{nn}$ $\sim$ 0.79 K. We found that the minimum eigenvalue of $V$ is at the Brillouin zone(BZ) center with $k$ = (0, 0), and the corresponding eigenvector represents a 120$^{\circ}$  state where the three spins in the unit cell lie in the plane pointing along three axes at angles 2$\pi/$3 to each other (Fig.2 (j)). Here, the large dipolar term breaks the rotation invariance, lifts the frustration of a kagome lattice and helps defeat the Mermin-Wagner theorem that forbids magnetic ordering in a 2D Heisenberg lattice. It is known that higher values of spin than 1/2 releases the frustration somewhat like soft spins would \cite{PRB1993}, and the case here has S = 7/2. This seems to enable a 2D-Ising like transition with a logarithmic heat capacity \`{a} la Onsager. Actually, similar spin structure was predicted by Maksymenko et. al by considering classical dipoles on a kagome lattice\cite{PRB2015Dipoles}. Their calculated specific heat actually agrees well with our experiment in the critical region by proper scaling (Fig. 2(c)). Thus, we conclude Mg$_2$Gd$_3$Sb$_3$O$_{14}$ to be a rare example of dipolar interaction mandated spin ordering on a Kagome lattice. 
	
	The Dy$^{3+}$ ion is an effective spin-1/2 Kramers ion with Ising anisotropy in an eight-oxygen-surrounding environment. Similar to the spin ice system, the Ising axis in Dy-TKL variant is along the lines joining each Dy to O1 (dashed lines in Fig. 1(c)). For the three sites in our Bravais lattice, the Ising directions are $\vec{\eta}_{blue}$ = $\frac{1}{\sqrt{5}} \{\sqrt{3}, 1, 1\}$, $\vec{\eta}_{red}$ = $\frac{1}{\sqrt{5}} \{-\sqrt{3}, 1, 1\}$, $\vec{\eta}_{green}$ = $\frac{1}{\sqrt{5}} \{0, -2, 1\}$ in the global cartesian frame. The small $\theta_{W}$ of -0.18 K corresponds to $J_{ex}$ $\sim$ 1.12 K, while the dipolar energy scale $D_{nn}$ $\sim$ 1.31 K is the largest energy scale. It is known that ferromagnetic spins with tripod-like anisotropy on a kagome lattice are highly frustrated, which will lead to the KSI state \cite{PRB2002Wills}. Similar to that of the pyrochlore spin ice \cite{Nature1999,Science2001}, the  ice rule (spins with either two-in-one-out or one-in-two-out respect to the center of each triangular) of KSI  will also result in great number of ground state degeneracy and zero-point entropy \cite{PRB2002Wills}. 
	
	With the TKL and strong dipolar interaction, it is tempting to view Mg$_2$Dy$_3$Sb$_3$O$_{14}$ as a realization of a dipolar ferromagnet where the KSI physics could be realized. 
	Our Luttinger-Tisza method yields a metastable state at the BZ center, which is an ordered KSI state. The corresponding spin structure (Fig. 2(k)) can be viewed as a three-sublattices ferromagnetic order with $k$ = 0.  This spin structure also resembles the theoretically predicted LRO state \cite{PRB2001SS,PRL1999SS,PRL2001MG} for the 3D pyrochlore spin ice observed in Tb$_2$Sn$_2$O$_7$ \cite{TbSnO}. However, this $k$ = 0  state is NOT a global ground state. The lowest eigenvalue of the exchange matrix is found to be at the six K points of BZ corners, whose energy is somewhat lower than that of the zone center. Such an eigenvalue corresponds to a LRO state with a 3$\times$3 tripled magnetic unit cell.  In addition, the magnitude of the moment differs in space, as prescribed by a commensurate spin density wave (SDW) state. Unlike the $k$ = 0 state where the KSI ice rule is preserved for every Mg-Dy tetrahedron (grey triangular in Fig. 2(k)), here one out of six tetrahedrons violates the local ice rule.   
	Note that the Luttinger-Tisza method used here, is more akin to the mean field theory when applied away from the zone center or the M point of the BZ. This is a nontrivial and complex problem (see e.g. Ref. \cite{SDW1,SDW2}), and requires further theoretical investigation. Regardless the exact nature of ordering, our TKL system then appears to enable the spin dynamics to be much more efficient, as compared to the 3D Dy$_2$Ti$_2$O$_7$ compound.  This interesting  contrast  therefore provides a strong impetus to the study of the underlying dynamics. Along with the observed LRO at 0.37 K, \textit{the Dy-Ising-TKL might provide a rare example exhibiting a LRO state that breaks the KSI degeneracy}.
	
	The Er$^{3+}$ ion in Mg$_2$Er$_3$Sb$_3$O$_{14}$ has a large angular momentum J = 15/2. At low temperatures, it reduces to an effective spin-1/2 as a result of the Kramers-doublet. The high temperature  $\theta_{W}$ $\sim$ -14.52 K implies a large exchange energy $J_{ex}$ $\sim$ 11.0 K, while the dipolar energy scale  is $D_{nn}$ $\sim$ 0.11 K by assuming a moment of 3 $\mu_B$ (similar to that in the Er pyrochlore). This implies that below $\sim$ 10 K the spins are locked up into the state preferred by the exchange. The single ion anisotropy term in this case gives rise to a local-XY model, where the Er$^{3+}$ are energetically favorable to lie in the local XY-plane perpendicular to the Ising axis discussed above. In the Er-pyrochlore, such a XY model, will give rise to a U(1) degeneracy in the spin Hamiltonian at the mean-field level that allows the Er$^{3+}$ spins to rotate continuously in the XY plane \cite{ErTiLB}.  In Er-TKL, similar XY degeneracy is preserved for the exchange part of the Hamiltonian. However, an arbitrarily small long range dipolar interaction will break the degeneracy. By diagonalizing the interacting matrix, a lowest energy eigenvalue is found at the BZ center whose eigenvector, by a curious coincidence, corresponds to the coplanar model exactly same as that of the Gd compound (Fig. 2(j)).
	
	Regarding the experimental observations for the Er-TKL, since the 2.1 K anomaly in $C(T)$ is extremely weak in terms of the entropy under the peak, and $\chi_{dc}$ shows a similarly weak anomaly, the order parameter might be one that still allows significant fluctuations below its $T_C$, reminiscent of a KT transition \cite{KT} where spin-vortices form and bind. Then the 80 mK transition shown on $\chi_{ac}$ is likely related to the predicted 120$^{\circ}$ coplanar AFM ordering. If so, such a low ordering temperature (frustration index $f$ = $\frac{\theta_{W}}{T_N}$ $\sim$ 180) suggests the importance of quantum spin fluctuations in terms of suppressing the ordering temperature and selecting the ordered state. The weak frequency dependence on $\chi_{ac}$ around the peak might indicate an increasing spin-lattice relaxation time as temperature is decreased. The importance of thermal coupling between a coherent spin system and the lattice needs to be understood for both identifying and potentially using quantum materials \cite{PNAS}.
	Future experiments including neutron scattering and Muon spin spectroscopy will be useful to identify the nature of the two transitions at 80 mK and 2.1 K.

	\textit{Summary.---}%
	We discovered  a new 2D rare earth TKL Mg$_2$RE$_3$Sb$_3$O$_{14}$ by partially substituting the ions in the cubic pyrochlore lattice. Our studies on three samples with RE = Gd, Dy, Er have already related their magnetism to various exotic states including the dipolar spin order, the KSI, and the KT transition. Due to the large variability of the spin sets in the rare earth family and the possibility of tuning the lattice parameters via chemical pressures, other exotic physics might also be realized. The future exploration of the whole TKL family members is expected to open a new field in condensed matter physics and materials science studies for coming years, such as the pyrochlore did during the last two decades.

	\begin{acknowledgments}
		{ The work of B.S.S. at UCSC was supported by the U.S. Department of Energy (DOE), Office of Science, Basic Energy Sciences (BES) under Award No. FG02-06ER46319. K.L. and Y.X.W. thank the support of the National Natural Science Foundation of China (grant No. 11275012). A.P.R was supported by NSF-DMR 1534741, and J.T. was supported by NSF DGE-1339067. The work at  NHMFL is supported by NSF-DMR-1157490 and State of Florida and the DOE and by the additional funding from NHMFL User Collaboration.}
		
	\end{acknowledgments}

\newpage

{\huge Supplemental material}
\section{1. Sample synthesis method and measurement setups}
Polycrystalline samples of tripod kagome lattice (TKL) compounds Mg$_2$RE$_3$Sb$_3$O$_{14}$ (RE = Gd, Dy, Er) were synthesized by  solid state reactions. Stoichiometric ratios of RE$_2$O$_3$ (RE = Gd, Dy, Er), MgO, and Sb$_2$O$_3$ powder were carefully ground and reacted  at a temperature of 1573 K for 60 hours with several intermediate grindings. Performing the reaction at temperatures above 1623 K will introduce site-disorder between the RE site and the Mg site, which is evidenced by the cubic pyrochlore phase in X-ray diffraction (XRD) patterns. The room temperature XRD patterns were measured with a HUBER X-ray powder diffractometer with the structural refinements performed using software package \textit{Fullprof-suite}. The dc susceptibility measurements were performed using a commercial superconducting interference device (SQUID) magnetometer with a magnetic field of 200 Oe. The ac susceptibility was measured at National High Magnetic Field Laboratory with the conventional mutual inductance technique at frequencies between 80 Hz and 700 Hz. The low temperature specific heat measurements were performed in a He3-He4 dilution refrigerator using the semi-adiabatic heat pulse technique. The powder samples were cold-sintered with Ag powder, the contribution of which was measured separately and subtracted from the data.  For all the specific heat data shown below, the magnetic contribution (C$_{mag}$) was obtained by subtracting a lattice contribution estimated from the results of a separate measurement of the non-magnetic isomorph Zn$_2$La$_3$Sb$_3$O$_{14}$.

\section{2. XRD pattern and Crystallographic table}
\begin{figure}[tbp]
	\linespread{1}
	\par
	\includegraphics[width = 3.4 in]{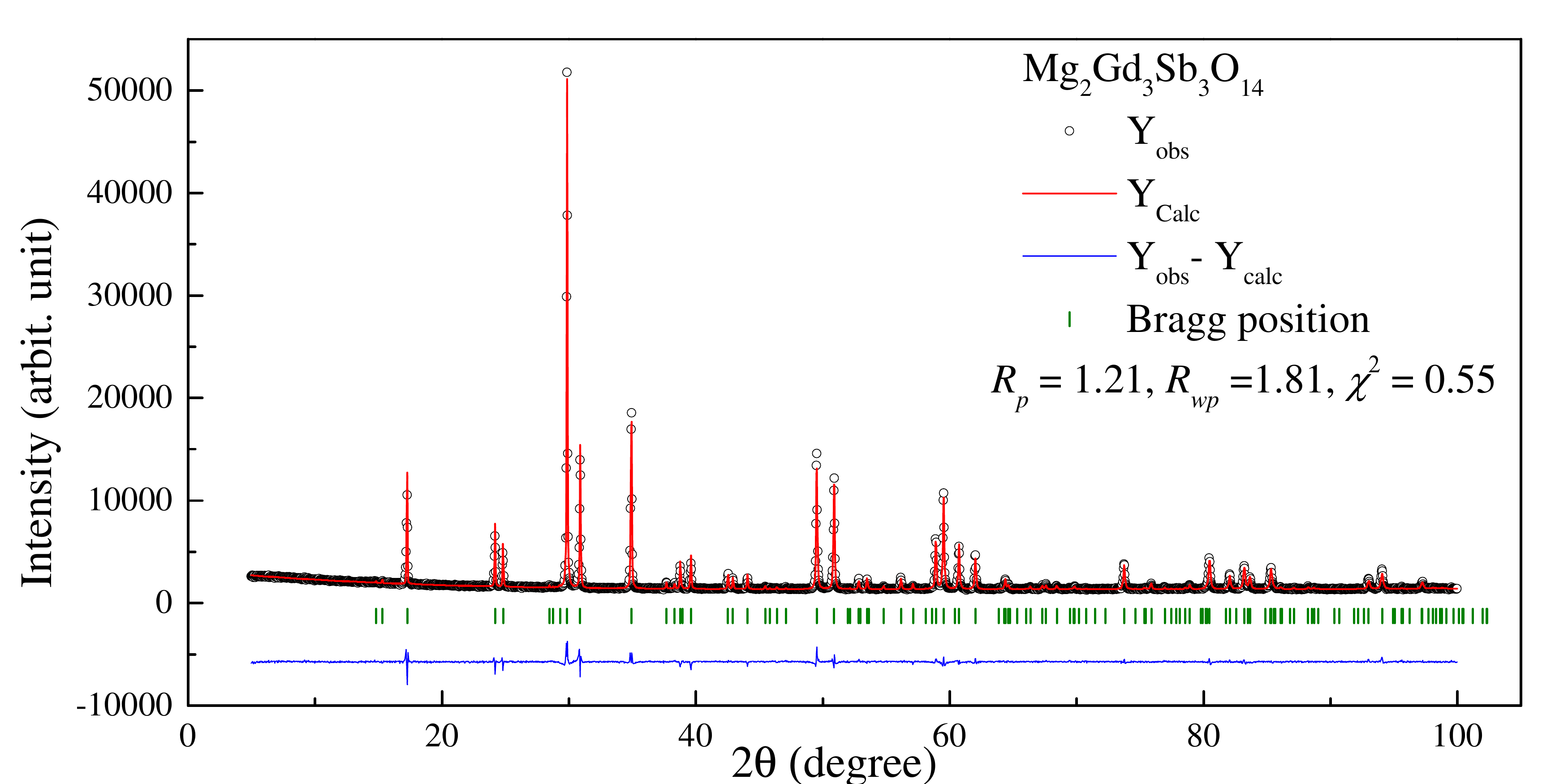}
	\par
	\caption{\label{Fig:S1}(Color online) The XRD pattern for Mg$_2$Gd$_3$Sb$_3$O$_{14}$. The solid curve is the best fits from
		the Rietveld refinement using \textit{Fullprof}. }
\end{figure}

The X-ray diffraction (XRD) pattern and Rietveld refinement result for Mg$_2$Gd$_3$Sb$_3$O$_{14}$ are shown in Fig.\ref{Fig:S1}. The XRD pattern for the other two compounds are similar to that of Mg$_2$Gd$_3$Sb$_3$O$_{14}$ except for the peak shifts due to the lattice parameter differences.  As expected, the lattice parameters decrease with the decreasing ionic size from Gd$^{3+}$ to Er$^{3+}$. The refined crystallographic results of atomic positions, lattice parameters, and selective bonds lengths are listed in Tab. \ref{Tab:1}.

\begin{table*} [tbp]
	\begin{center}
		\caption{ \label{Tab:1} Crystallographic parameters and selected bond lengths from Rietveld refinement of room temperature XRD patterns for Mg$_2$RE$_3$Sb$_3$O$_{14}$ (RE = Gd, Dy, Er).}
		\begin{tabular}{ccccc}
			\hline 
			RE	&   & Gd  & Dy  & Er \\ 
			\hline 
			IR (RE$^{3+}$) (\AA) &  & 1.08 & 1.05  & 1.03  \\
			a (\AA) &  & 7.35644(3) & 7.31865(8) &   7.29226(10) \\ 
			c (\AA) &  &  17.35214(9) & 17.2868(2)   & 17.2365(3)\\ 
			\hline 
			Atom & Wyckoff site & $x, y, z$  & $x, y, z$  & $x, y, z$ \\
			\hline 
			Mg1      & 3a  & 0, 0, 0    & 0, 0, 0  & 0 ,0 ,0 \\ 
			Mg2      & 3b & 0, 0, 1/2   & 0, 0, 1/2 & 0, 0, 1/2  \\ 
			RE      & 9d  & 1/2, 0, 1/2 & 1/2, 0, 1/2 & 1/2, 0, 1/2 \\ 
			Sb      & 9e  & 1/2, 0, 0 & 1/2, 0, 0 &  1/2, 0, 0 \\
			O1      & 6c  & ~~0, 0, 0.1027(10)~~ & ~~0, 0, 0.1038(10)~~ &  ~~0, 0, 0.1034(16) \\ 
			
			O2$_x$  &  & 0.5361(7) & 0.5411(7) &  0.5377(12) \\ 
			O2$_y$ &  18h & 0.4639(7) & 0.4589(7) &   0.4623(12) \\ 
			O2$_z$   &  &  0.8876(6) & 0.8850(6) &  0.8883(9) \\ 
			
			O3$_x$   &   & 0.4752(7) & 0.4797(7) &  0.4793(12) \\ 
			O3$_y$  & 18h & 0.5248(7) & 0.5203(7) &  0.5207(12) \\ 
			O3$_z$   &  & 0.3590(5) & 0.3608(5) &  0.3624(9) \\ 
			\hline 
			RE-RE (intra-plane) (\AA)		 & & 3.67822(1) & 3.65934(5) &  3.64613(3) \\
			RE-RE (inter-plane) (\AA)		 & & 6.16157(3) & 6.13739(6) &  6.11900(10) \\
			RE-O1 (\AA)      & & 2.395(9) & 2.375(8)  &    2.370(14)        \\
			RE-O2 (\AA)      & & 2.566(5) & 2.576(6) &   2.562(8)          \\
			RE-O3 (\AA)      & & 2.467(9) & 2.420(11) &   2.386(16)          \\
			B (${\AA}^2$) &   &  1.32(2) & 1.37(3) &  1.47(4)   \\ 
			R$_p$		&	&  1.21   &  1.60 &   3.73  \\
			R$_{wp}$ 	&	&  1.81   &  2.33 &    5.32  \\
			${\chi}^2$  &  &  0.554 &  0.939 &   2.81   \\ 
			\hline
		\end{tabular} 
	\end{center}
\end{table*}

\section{3. Magnetization and single ion anisotropy}
\begin{figure*}[t]
	\linespread{1}
	\par
	\includegraphics[width = 6.5 in]{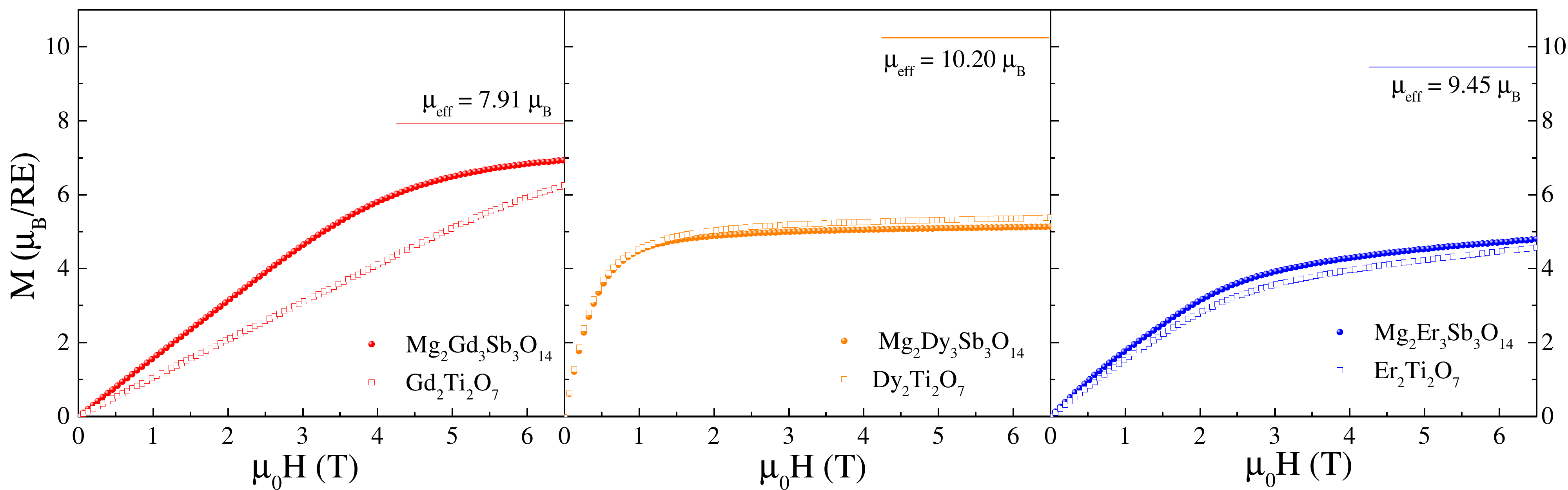}
	\par
	\caption{\label{Fig:S2}(Color online) Magnetization curves up to 6.5 T measured at $T$ = 2 K for Mg$_2$RE$_3$Sb$_3$O$_{14}$ and RE$_2$Ti$_2$O$_7$ (RE = Gd, Dy, Er). The $\mu_{eff}$ is the effective moment of RE ion for each RE-TKL found by the Curie-Weiss fit of the dc susceptibility.}
\end{figure*}

The magnetization curves at 2 K for three TKL compounds and their pyrochlore cousins ( all in polycrystalline forms) are shown shown in Fig. \ref{Fig:S2}, where apparent similarities are found for all three sample sets.

For Gd-TKL, the magnetization curve shows a straight line up to 3 T, which signatures a isotropic g-factor as expected for a S state of Gd$^{3+}$ (J = 7/2, L = 0). The magnetization reaches 7.0 $\mu_B$/Gd$^{3+}$ (6.4 $\mu_B$/Gd$^{3+}$ for Gd$_2$Ti$_2$O$_7$)  at 6.5 T, which is about the 88.5\% of its effective moment.

Similar to Dy$_2$Ti$_2$O$_7$, strong anisotropic behaviors are observed in Dy-TKL that the magnetization quickly saturates at a plateau of 5.1 $\mu_B$/ Dy$^{3+}$, which is about half of its effective moment. In the pyrochlore spin ice, such half magnetization plateau is a characteristic behavior of Dy$^{3+}$ Ising moment \cite{DTO}. Similar half magnetization plateau observed here provides a strong evidence for the local Ising anisotrpy for Dy ion in the TKL system.

For Er-TKL, the magnetization reaches 4.8 $\mu_B$/Er$^{3+}$ (4.6 $\mu_B$/Er$^{3+}$ for Er$_2$Ti$_2$O$_7$), which is about half of the effective moment per Er ion. Moreover, the magnetization curve of Er-TKL is also very similar to that of its pyrochlore cousin Er$_2$Ti$_2$O$_7$ for the whole field region from 0 to 6.5 T. This similarity provides a supporting evidence for the local XY anisotropy that has been confirmed in the Er pyrochlore.

\section{4. Model and parameters}
For the kagome lattice described in the main text, ${\bf A_1}$ and ${\bf A_2}$ are basis vectors of the triangular Bravais lattice where there are three basis sites in a unit cell for an upright triangle of basis spins, labeled as blue, red and green. The nearest distance between two spin are thus $R_{nn}$ = $\frac{1}{2}A_1$. The positions of the Bravais lattice points are denoted by ${\bf R}^{l}$, whereas each of the three sites in the unit cell is labeled by ${\bf r}_{a}$ (where a = blue, red, green). So each site can be labled as  ${\bf R}_a^{l}$ =  ${\bf R}^{l}$ + ${\bf r}_{a}$. 

The general Hamiltonian of the system is:
\begin{eqnarray}
	\label{eq:H2}
	H=\frac{1}{2}\sum_{k,l,a,b} \sum_{\alpha,\beta}V_{ab}^{\alpha\beta}\left({\bf R}_{ab}^{kl}\right)S_a^{\alpha}\left({\bf R}^{k}\right)S_b^{\beta}\left({\bf R}^{l}\right) \nonumber\\	
	 +\sum_{l,a,\alpha}C\left(S^{\alpha}\left({\bf R}^l_a\right){\eta}^{\alpha}\left({\bf R}^l_a\right)\right)^2
\end{eqnarray}
\begin{equation}
	\label{eq:H3}
	V_{ab}^{\alpha\beta}\left(\bf R\right)=D_{nn}R_{nn}^3\left(\frac{\delta_{\alpha\beta}}{|{\bf R}|^3}-3\frac{R_{\alpha}R_{\beta}}{|{\bf R}|^5}\right)+J_{ex}\delta_{\alpha\beta}^{|{\bf R}=R_{nn}|}
\end{equation}
\begin{figure*}[tbp]
	\linespread{1}
	\par
	\includegraphics[width = 6.5 in]{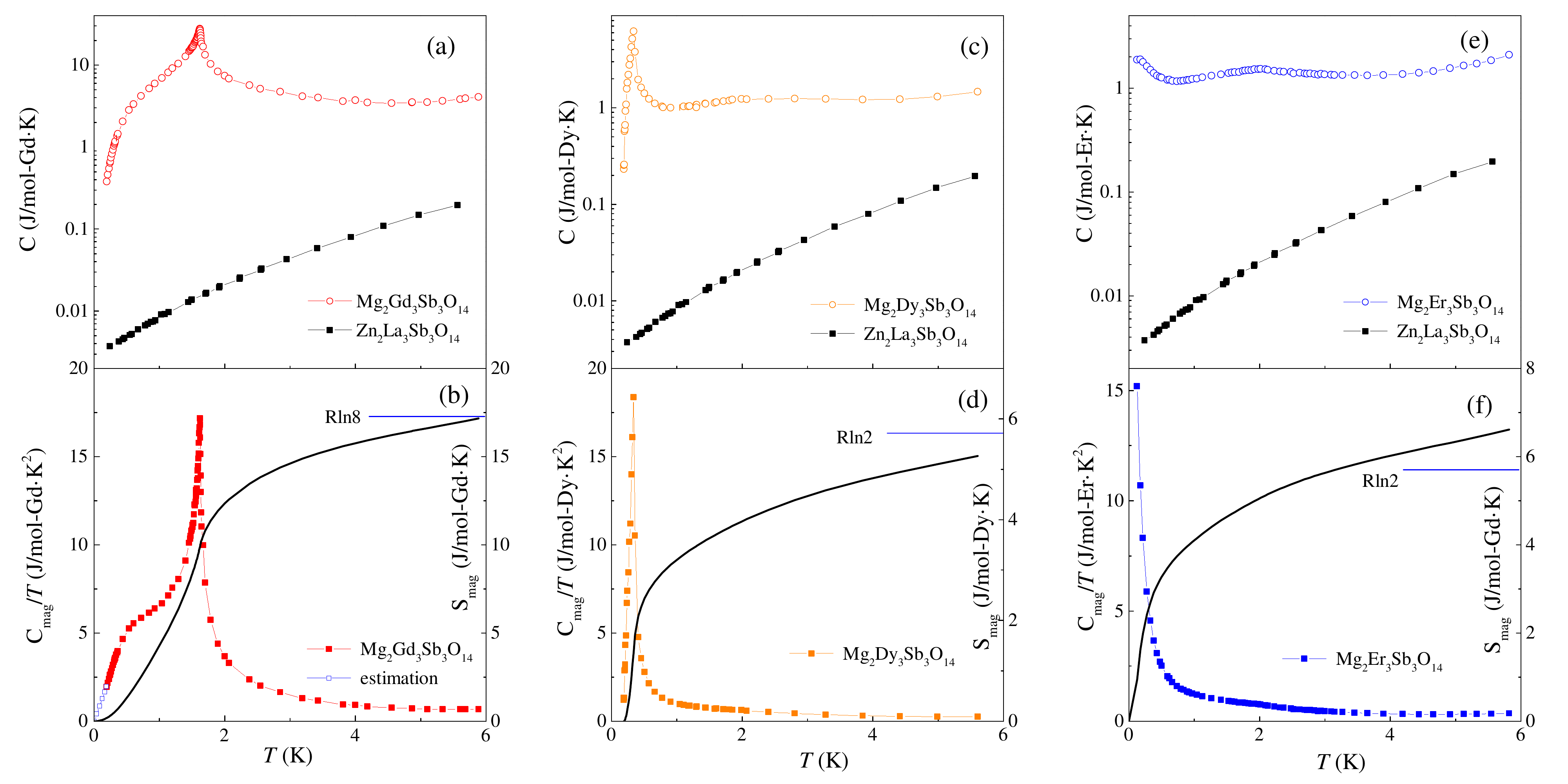}
	\par
	\caption{\label{Fig:S3}(Color online) (a,c,e) Measured total specific heat of three TKL compounds and the lattice contribution measured with Zn$_2$La$_3$Sb$_3$O$_{14}$.  (b,e,f) Magnetic specific heat over temperature and the integrated magnetic entropy for Mg$_2$RE$_3$Sb$_3$O$_{14}$ (RE = Gd, Dy, Er).}
\end{figure*}
The ${\bf R}_{ab}^{kl}$ is the vector between two classical spins $S_a^{\alpha}\left({\bf R}^{k}\right)$, $S_b^{\beta}\left({\bf R}^{l}\right)$ of the unit length. $k$ and $l$ index the unit cell, $a$ and $b$ run over the sites in the unit cell, while $\alpha$ and $\beta$ denote the x, y, z components of the spin vectors. The first term in Eq.(2) is the interacting term while the second term is the single ion term. Fourier transformation of the Equation (2) yields the Hamiltonian in the wave vector space that is listed in the main text. ${\bm {\eta}}({\bf R}_a^l)$ represents the local Ising axis of site ${\bf R}_a^l$ as given in the main text. To enforce the spin anisotropies (e.g Ising-like for Dy ion and local XY-like for Er ion), the sign of coefficient $C$ is taken  negative for Dy ion and positive for Er ion while the magnitude of $C$ is chosen ten times larger than any other energy scale. $C$ is zero for the half filled spin case of Gd.

The first term in $V_{ab}^{\alpha\beta}$ represents dipolar interaction and $D_{nn}$ is the dipolar energy scale that takes the form: 
\begin{equation}
	D_{nn} = \frac{\mu_0}{4\pi}\frac{\mu^2_{eff}}{R^3_{nn}}
\end{equation}
where $\mu_{eff}$ is the effective moment of RE ions. For each RE-TKL, $R_{nn}$ is found by the Retvield refinement of the XRD pattern and listed in Table.1.

The second term in matrix $V_{ab}^{\alpha\beta}$ represents the nearest neighbor exchange interaction. The effective exchange constant $J_{ex}$ is positive for antiferromagnetic exchange coupling and negative for ferromagnetic exchange coupling. An estimation of the exchange constant $J$ can be found from the measured Weiss temperature $\theta_W$ via a mean field theory \cite{Kittel}:
\begin{equation}
	J = -\frac{3\theta_W}{zS(S+1)}
\end{equation}
where $z$ is the number of nearest neighbor. In our TKL system, $z$ = 4. Since $J_{ex}$ is the effective exchange constant that couples two unit spins, $J_{ex} = JS(S+1)$.

Because of the large spin moment in the RE-oxide systems, $\theta_W$ has the contribution from both the exchange and dipolar parts. As a simple approximation, one can estimate the  exchange contribution by subtracting $\theta_W$ by the dipolar contribution $D_{nn}$ \cite{CW}.
Thus we get:
\begin{equation}
	J_{ex} = -\frac{3}{4}(\theta_W-D_{nn})
\end{equation}
Adopting the parameters for our TKL system, we get (1) $D_{nn}$ = 0.79 K, $J_{ex}$ = 6.1 K for Gd compound, (2) $D_{nn}$ = 1.31 K, $J_{ex}$ = 1.12 K for Dy compound, and (3) $D_{nn}$ = 0.11 K, $J_{ex}$ = 11.0 K for Er compound.

In order to determine their ground states, we have used a Luttinger-Tisza type theory to calculate the eigenvalues and eigenfunctions of the interaction matrix, following two early references by Luttinger-Tisza \cite{LT1} and Onsager \cite{LT2}. These fundamental papers outline a procedure for simple cubic cell with a single magnetic species.  Their idea  is adapted for the case of Kagome lattice, viewed as having  three independent spins in a triangular  unit cell. We compute the 9$\times$9 interaction matrix of $V$ at each wave vector numerically, and search for the absolute minimum eigenvalue.  By scanning all wave-vectors of the Brillouin Zone (BZ), typically the minima lie on high symmetry lines, we evaluate all possible  periodicities of the magnetic state. The eigenvectors are decomposed to find the  magnitude of moments on each of the three spins, if these are normalizable to the same constant we have an acceptable solution. The  case of  a degenerate  eigenvalue is subtle and often comes up in our analysis. In this case, we perform a test: if   no linear combination of the eigenvectors can be found leading  to an equal normalization of the three spins, then the solution is unphysical and must be discarded. Whenever the minimizing wave vector is different from $\Gamma$ or M point of the BZ, the solution found by this method is analogous to a mean field spin density wave. Since the magnitudes of the spins, which is given by the real part of the wave, must vary in magnitude at different sites, this does not fulfill the conditions of Luttinger and Tisza.

\section{5. Magnetic specific heat and Entropy}

For all the TKL system, the magnetic specific heat (C$_{mag}$) was obtained by subtracting the total specific heat (C) by a lattice contribution estimated from the results of a separate measurement of the non-magnetic isomorph Zn$_2$La$_3$Sb$_3$O$_{14}$. Below 6 K, the lattice contribution is almost two magnitude smaller than the magnetic contribution as shown in Fig. \ref{Fig:S3}(a,c,e). The corresponding integrated magnetic entropies (S$_{mag}$) are shown in Fig. \ref{Fig:S3}(b,d,f). For Mg$_2$Er$_3$Sb$_3$O$_{14}$,  the integrated S$_{mag}$ from 0.11 K to 6 K reaches 6.61 J/mol-Er$\cdot$K. Due to the possible magnetic ordering at 80 mK and possible contribution from the $\textsuperscript{167}$Er nuclear spin, it is unlikely that the integrated entropy has any significance.

\end{document}